\documentstyle[aps,prl,twocolumn,psfig]{revtex}
\begin{document}
\title{ The Dynamic Exponent of the Two-Dimensional Ising Model and Monte Carlo
Computation of the Sub-Dominant Eigenvalue of the Stochastic Matrix }
\author{ M. P. Nightingale} 
\address{ Department of Physics, University of Rhode Island, Kingston RI
02881, USA}
\author{H.W.J. Bl\"{o}te }
\address { Department of Applied Physics, Delft University of Technology, 
Lorentzweg 1, 2628 CJ Delft, The Netherlands }
\date{\today}
\maketitle
\begin{abstract}
We introduce a novel variance-reducing Monte Carlo algorithm
for accurate determination of autocorrelation times.
We apply this method to two-dimensional Ising systems 
with sizes up to $15 \times 15$, using single-spin flip dynamics,
random site selection and transition probabilities according to the
heat-bath method. 
{}From a finite-size scaling analysis of these autocorrelation times,
the dynamic critical exponent $z$ is determined
as $z=2.1665$ (12).
\end{abstract}
\pacs{64.60.Ht, 02.70.Lq, 05.70.Jk, 64.60.Fr}

The onset of criticality is marked by a divergence of both the
correlation length $\xi$ and the correlation time $\tau$.  While the
former divergence yields singularities in static quantities, the latter
manifests itself notably as critical slowing down.  To describe dynamic
scaling properties, only one exponent is required in addition to the
static exponents.  This dynamic exponent $z$ links the divergences of
length and time scales: $\tau \sim \xi^z$. In our computation of $z$ we
exploit that for a finite system $\xi$ is limited by the system size
$L$, so that $\tau \sim L^z$ at the incipient critical point.

In this Letter, we focus on the two-dimensional Ising model with
Glauber-like dynamics.  Values quoted in the literature for $z$ vary
vastly, from $z=1.7$ to $z=2.7$~\cite{z-range}, but recent computations
seem to be converging towards the value reported here.  Finally,
results are beginning to emerge of precision sufficient for sensitive
tests of fundamental issues such as universality.

The numerical method introduced in this Letter is related to Monte
Carlo methods used to compute eigenvalues of Hamiltonians of discrete
or continuous quantum systems\cite{Zhangetc95,Taka} and transfer
matrices of statistical mechanical systems\cite{NighGraKos}.  In
particular, the current method is suitable to obtain more than one
eigenvalue by adaptation of the diffusion Monte Carlo algorithm of
Ref.~\onlinecite{CeperleyBernu88+BernuCeperley90}.

To compute the autocorrelation time of small $L \times L$ lattices we
exploit the following properties of the single-spin-flip Markov (or
stochastic) matrix $\bf P$~\cite{markov}.  It operates in the linear
space of all spin configurations and its largest eigenvalue 
equals unity.  The corresponding right eigenvector contains the
Boltzmann weights of the spin configurations; the left eigenvector is 
constant, reflecting probability conservation.  The correlation
time $\tau_L$ (in units of one flip per spin, i.e., $L^2$ single-spin
flips) is determined by the second-largest eigenvalue $\lambda_L$:
\begin{equation}
\tau_L = -\frac{1}{ L^2 \ln \lambda_L}.
\label{taulab}
\end{equation}
For a system symmetric under spin inversion, the corresponding
eigenvector is expected to be antisymmetric.

We used two methods to compute $\lambda_L$:  exact, numerical
computation for $L\le 5$, and Monte Carlo
for $4 \leq L \leq 15$.  The exact method used the conjugate
gradient algorithm~\cite{NiViMuel} and the symmetries of periodic
systems.  This calculation resembles that in
Ref.~\onlinecite{NBPhysica}, but currently we realize Glauber-like
dynamics using heat-bath or Yang~\cite{Yang} transition
probabilities and random site selection.

The Monte Carlo method used a stochastic form of the power method,
as follows.  A spin configuration $s$ with energy $E(s)$ 
has a probability
\begin{equation}
\frac{\exp(-E(s)/kT)}{Z} \equiv \frac{ \psi_{\rm B}(s)^2}{Z},
\end{equation}
where $Z$ is the partition function.
The element $P(s'|s)$ of the Markov matrix is the probability
of a single-spin-flip transition from 
$s$ to $s'$.  Since $\bf P$ satisfies detailed balance, 
\begin{equation}
\hat P(s'|s)\equiv {1 \over \psi_{\rm B}(s')} P(s'|s) \psi_{\rm B}(s)
\end{equation}
is symmetric.
For an arbitrary trial state $|f\rangle$ an effective eigenvalue
$\lambda^{(t)}_L$ is defined by
\begin{equation}
\lambda^{(t)}_L =
{\langle {\bf \hat P}^{t+1} \rangle_f \over
 \langle {\bf \hat P}^t\rangle_f}, 
\label{eq.lambdat}
\end{equation}
where $\langle \cdot \rangle_f$ is the expectation value in the state
$|f\rangle$.  In the limit $t \to \infty$ the effective eigenvalue
converges generically to the dominant eigenvalue allowed 
by the symmetry of $|f\rangle$. The convergence is 
exponential in the time lag $t$.

Given a trial state $|f\rangle$, standard Monte Carlo suffices to 
compute the right-hand side of Eq.~(\ref{eq.lambdat}), i.e.,
the denominator of Eq.~(\ref{eq.lambdat}) 
\begin{displaymath}
N^{(t)} \equiv \langle f |{\bf \hat P}^t| f \rangle =\nonumber \\
\end{displaymath}
\begin{displaymath}
\sum_{s_1, \dots, s_{t+1}} f(s_{t+1}) \hat P(s_{t+1}1|s_t)\cdots
\hat P(s_2|s_1) f(s_1)=\nonumber \\
\end{displaymath}
\begin{displaymath}
\sum_{s_1, \dots, s_{t+1}} {f(s_1)f(s_{t+1}) \over 
\psi_{\rm B}(s_1)\psi_{\rm B}(s_{t+1})}
P(s_{t+1}|s_t)\cdots P(s_2|s_1)
\psi_{\rm B}(s_1)^2=\nonumber \\
\end{displaymath}
\begin{equation}
Z \left \langle {f(s_1)f(s_{t+1}) \over \psi_{\rm B}(s_1)\psi_{\rm B}(s_{t+1})}
\right \rangle_P,
\label{eq.N}
\end{equation}
is an autocorrelation;
$f(s)\equiv \langle s|f\rangle$ and $\langle \cdot \rangle_P$ denotes 
the average with respect to the probability
\begin{equation}
P(s_{t+1}|s_t)\cdots P(s_2|s_1) \psi_{\rm B}(s_1)^2/Z
\label{prdis}
\end{equation}
of finding a configuration $s_1$ in equilibrium and subsequent
transitions to configurations $s_2$ through $s_{t+1}$.

Similarly, the numerator of Eq.~(\ref{eq.lambdat}) 
\begin{eqnarray}
\label{eq.H}
H^{(t)} \equiv \langle f |\hat P^{t+1}| f \rangle = \nonumber \\
\sum_{s_0, \dots, s_{t+1}} f(s_{t+1}) \hat P(s_{t+1}|s_{t})\cdots
\hat P(s_1|s_0) f(s_0)= \nonumber \\
{1 \over 2}Z \left \langle [\lambda_L(s_1)+\lambda_L(s_{t+1})]
{f(s_1)f(s_{t+1}) \over \psi_{\rm B}(s_1)\psi_{\rm B}(s_{t+1})}
\right
\rangle_P
\end{eqnarray}
is a cross-correlation,
where the `configurational eigenvalue' $\lambda_L(s)$ of
spin configuration $s$ is defined as 
\begin{equation}
\lambda_L(s) = {1 \over f(s)} \sum_{s'} f(s') \hat P(s'|s).
\label{eq.lamdas}
\end{equation}
Finally, with Eqs.~(\ref{eq.N}) and (\ref{eq.H}), one has
$\lambda_L^{(t)} = H^{(t)}/N^{(t)}$ for the effective
eigenvalue. 

In practice, $H^{(t)}$ and $N^{(t)}$ are estimated by conventional Monte
Carlo methods. As usual, these estimators involve time averaging of 
stochastic variables. Thus, on the right of Eqs.~(\ref{eq.N}) and 
(\ref{eq.H}) $s_i$ is replaced by $s_{t'+i-1}$ ($i=1,\dots,t$), and the 
Monte Carlo average is taken over an appropriately chosen subset of 
times $t'$ after thermal equilibration.

In principle, one could choose $f=m \psi_{\rm B}$,  where $m$ is the
magnetization.  In that case, the above method reduces to estimating
the effective eigenvalue of the Markov matrix in terms of the
magnetization autocorrelation function $g(t)$ via
$\lambda_L^{(t)}=g(t+1)/g(t)$.  To estimate $g(t)$ one would average
over time products of the form $m(s_1)m(s_{t+1})$.  Eq.~(\ref{eq.H}) would
then yield $g(t+1)$ by replacing $m(s_t)$ by the conditional
expectation value of the magnetization at time $t+1$, evaluated
explicitly as $\sum_{s_{t+1}} m(s_{t+1})P(s_{t+1}|s_t)$.

The crux is that the estimator of $\lambda_L^{(t)}$ satisfies a
zero-variance principle, since Eqs.~(\ref{eq.N}) and (\ref{eq.H})
contain an optimizable trial state $|f\rangle$.  In the ideal case,
$|f\rangle$ is an exact eigenstate of the symmetrized Markov matrix
$\bf \hat P$, and the `configurational eigenvalue' $\lambda_L(s)$
equals the eigenvalue independent of $s$.  Then, the estimator of the
effective eigenvalue $\lambda_L^{(t)}$ yields the exact eigenvalue
without statistical and systematic errors at finite $t$, if care is
taken to arrange cancellation of the fluctuating factors in the
estimators of $H^{(t)}$ and $N^{(t)}$.  In practice, $|f\rangle$ is not
an exact eigenstate, and this introduces statistical and systematic
errors. However, these errors are kept small by the zero-variance
principle, if the trial states are accurate.

Such optimized trial states are constructed prior to the main Monte
Carlo run, by minimization of the variance $\chi^2$ of the
configurational eigenvalue
\begin{equation}
\chi^2(p)=
{\langle ({\bf \hat P} -\langle {\bf \hat P} \rangle_f})^2\rangle_f.
\end{equation}
As indicated, the variance depends
on the parameters $p$ of the trial state.  Optimization over $p$ is
done following Umrigar {\it et al.} \cite{cyrus.trialfunctions}:
one samples $M$ configurations $s_i$, typically a few thousand, with
probability ${\psi_{\rm B}^2 Z^{-1}}$ and approximates $\chi^2(p)$ by
\begin{equation}
\label{eq.chi_app}
\chi^2(p) \approx {\sum_{i=1}^M [{f(s_i,p) /
\psi_{\rm B}(s_i)}]^2 [\lambda_L(s_i,p)-\bar{\lambda}_L(p)]^2 \over 
\sum_{i=1}^M {[f(s_i,p) / \psi_{\rm B}(s_i)]^2 }}.
\end{equation}
Here $\bar{\lambda}_L$ denotes the weighted average of the
configurational eigenvalue over the sample, while the modified notation
explicitly shows dependences on the parameters $p$ of the trial state
$|f\rangle$.  Near-optimal values of the parameters $p$ can be obtained
by minimization of the expression on the right-hand side of
Eq.~(\ref{eq.chi_app}) for a {\it fixed} sample.

Analysis of the exact left eigenvectors of the Markov matrix $\bf P$
for systems with $L \leq 5$ shows that the elements depend only on the
magnetization to good approximation. This suggests trial functions
depending on long-wavelength components of the Fourier transform of
$s_i$, the zero-momentum component of which is just the magnetization
$m$.  The form
\begin{equation}
f(s) = \tilde{\psi}_{\rm B}(s) \; \psi^{{(+)}}(s) \; \psi^{{(-)}}(s),
\end{equation}
where $\psi^{(\pm)}\to \pm \psi^{(\pm)}$ under spin inversion, yields
an antisymmetric trial function, as required.  The tilde in
$\tilde{\psi}_{\rm B}$ indicates that the temperature is used as a
variational parameter, but we found that its optimal value 
is virtually indistinguishable from the true temperature.
The $\psi^{(\pm)}$ were chosen as
\begin{eqnarray}
\psi^{{(+)}}=\sum_{\bf k} a_{\bf k}(m^2) m_{\bf k}^{{(+)}} +
 m \sum_{\bf k} b_{\bf k}(m^2) m_{\bf k}^{{(-)}} \\
\psi^{{(-)}}=m \sum_{\bf k} c_{\bf k}(m^2) m_{\bf k}^{{(+)}} +
 \sum_{\bf k} d_{\bf k}(m^2) m_{\bf k}^{{(-)}},
\end{eqnarray}
where the index ${\bf k}$ runs through a small set of multiplets of
four or less long-wavelength wave vectors defining the $m_{\bf
k}^{(\pm)},$ translation and rotation symmetric sums of products of
Fourier transforms of the local magnetization; the ${\bf k}$ are
selected so that $m_{\bf k}^{(-)}$ is odd and $m_{\bf k}^{(+)}$ is even
under spin inversion; the coefficients $a_{\bf k},\;b_{\bf k},\;c_{\bf
k}$ and $d_{\bf k}$ are polynomials of second order or less in $m^2$.
The degrees of these polynomials were chosen so that no terms occur of
higher degree than four in the spin variables.  No more than forty
parameters were optimized for the trial functions used in the
computations reported here.

Since the probability distribution Eq.~(\ref{prdis}) is precisely the
one purportedly generated by standard Monte Carlo, the sampling 
procedure is straightforward. The Monte Carlo algorithm used a
random-number generator of the shift-register type. It was selected
with care to avoid the introduction of systematic errors; see
discussion and references in Ref.~\onlinecite{BLH}. We used two
Kirkpatrick-Stoll~\cite{KS}  generators, the results of which were
combined by a bitwise exclusive-or \cite {HBC}. For test purposes we
replaced one Kirkpatrick-Stoll generator by a linear congruential rule, 
but this did not reveal clear differences~\cite{BLH}.

For each system size $4 \leq L \leq 15$, Monte Carlo averages were
taken over $8 \times 10^8$ spin configurations. For $L=13$ - 15
these were separated by intervals of 16 sweeps (Monte Carlo steps per
spin); 8 sweeps for $L=11$ and 12; 2 sweeps for $L=5$ and 6; and only
one sweep for $L=4$.  The simulations of the remaining system sizes
consisted of parts using intervals of 2, 4 or 8 sweeps.

The numerical results for the effective second largest eigenvalue
$\lambda_L^{(t)}$ as a function of the projection time $t$ appeared to
converge rapidly.  In agreement with scaled results for $L \leq 5$
spectra, we observe that convergence occurs within a few intervals as
given above. Monte Carlo estimates of $\lambda_L$ are shown in Table
\ref{tab:numres}, as are exact results for small systems. For system
sizes $L=4$ and 5, the two types of calculation agree
satisfactorily.  The small numerical errors indicate that the
variance-reducing method introduced above is quite effective.

For finite system size $L$ there are corrections to the leading scaling
behavior $\tau_L \sim L^z$.  In the two-dimensional Ising model
corrections to static equilibrium quantities occur with even powers of
$1/L$ \cite{BdN}; thus we expect
\begin{equation}
\tau_L \approx L^z \sum_{k=0}^{n_{\rm c}} \alpha_k L^{-2k},
\label{fit}
\end{equation}
where the series was arbitrarily truncated at order $n_{\rm c}$, but
other powers of $1/L$ might occur as well. Ignoring the latter, we
fitted the autocorrelation times of Table \ref{tab:numres} to this
form.  Typical results of such fits are given in Table
\ref{tab:fitres}.  The smallest systems do not fit Eq.~(\ref{fit}) 
well, at least not for the $n_{\rm c}$ values used. The
residuals decrease rapidly when the minimum system size is increased
and the consistency between the results for different $n_{\rm c}$
suggests that Eq.~(\ref{fit}) captures the essential scaling behavior
of $\tau_L$.
{}From these results we chose the entry for $L \geq 5$ and $n_{\rm c}=2$ 
as our best estimate: $z=2.1665 \pm 0.0012$, where we conservatively quote
a two-sigma error. To our knowledge, this is the most precise estimate
of $z$ obtained to date, as evidenced by recent results summarized
in Table \ref{tab:compare}.  The table shows that that the mutual 
consistency of the results for $z$ has tended to improve in recent years.
The only recent result that appears inconsistent with ours is due to Li
{\it et al.}\cite{LiSZ}.  Its error is copied from  Table I of Li 
{\it et al.}. The data in this table display finite-size dependences that 
exceed the quoted errors, which may explain 
the discrepancy with our result.

This research was supported  by the (US) National Science Foundation
through Grant \# DMR-9214669, by the Office of Naval Research and by
the NATO through Grant \# CRG 910152.  This research was conducted in
part using the resources of the Cornell Theory Center, which receives
major funding from the National Science Foundation (NSF) and New York
State, with additional support from the Advanced Research Projects
Agency (ARPA), the National Center for Research Resources at the
National Institutes of Health (NIH), IBM Corporation, and other members
of the center's Corporate Research Institute.


\begin{table}[htbp]
\caption{Second-largest eigenvalue 
$\lambda_L$ of the Markov matrix. The first column 
indicates the method: exact numerical or Monte Carlo.}
\vskip 1 ex
\begin{center}
\begin{tabular}{||c|r|l|l||}
method&$L$&$\mbox{\hspace{15mm}}\lambda_L$&\mbox{\hspace{5mm}}error\\
\hline
exact & 2 &  0.985702260395516 & 0.000000000001 \\
exact & 3 &  0.997409385126011 & 0.000000000001 \\
exact & 4 &  0.999245567376453 & 0.000000000001 \\
exact & 5 &  0.999708953624452 & 0.000000000001 \\
\tableline
MC    & 4 &  0.9992455685      & 0.0000000094   \\
MC    & 5 &  0.9997089453      & 0.0000000060   \\
MC    & 6 &  0.9998657194      & 0.0000000045   \\
MC    & 7 &  0.9999299708      & 0.0000000031   \\
MC    & 8 &  0.9999600854      & 0.0000000023   \\
MC    & 9 &  0.9999756630      & 0.0000000017   \\
MC    &10 &  0.9999843577      & 0.0000000014   \\
MC    &11 &  0.9999895056      & 0.0000000010   \\
MC    &12 &  0.9999927107      & 0.0000000008   \\
MC    &13 &  0.9999947840      & 0.0000000006   \\
MC    &14 &  0.9999961736      & 0.0000000005   \\
MC    &15 &  0.9999971314      & 0.0000000005   \\
\end{tabular}
\end{center}
\label{tab:numres}
\end{table}
\pagebreak
\begin{table}[htbp]
\caption{ Results of least-squares fits for the dynamic exponent.
The first column shows the minimum system size included, the second 
the number of correction terms included, and the third column whether (y)
or not (n) numerical exact results (for $L\leq5$) are included. The
last column contains the chi-square confidence index.\protect\cite{NumR}.
}
\vskip 1 ex
\begin{center}
\begin{tabular}{||c|c|c|c|c|c||}
$L\geq$&$n_{\rm c}$&exact&$\mbox{\hspace{3mm}}z$&error&$Q$\\
\hline
    4 & 1 & n & 2.1769 &  0.0001   & 0.00 \\
    5 & 1 & n & 2.1705 &  0.0002   & 0.00 \\
    6 & 1 & n & 2.1688 &  0.0003   & 0.23 \\
    7 & 1 & n & 2.1679 &  0.0006   & 0.43 \\
    8 & 1 & n & 2.1672 &  0.0010   & 0.42 \\
\tableline
    4 & 2 & n & 2.1650 &  0.0003   & 0.17 \\
    5 & 2 & n & 2.1665 &  0.0006   & 0.70 \\
    6 & 2 & n & 2.1662 &  0.0013   & 0.60 \\
    7 & 2 & n & 2.1648 &  0.0024   & 0.52 \\
\tableline
    4 & 3 & n & 2.1672 &  0.0009   & 0.64 \\
    5 & 3 & n & 2.1656 &  0.0020   & 0.61 \\
    6 & 3 & n & 2.1625 &  0.0044   & 0.56 \\
    3 & 3 & y & 2.1653 &  0.0004   & 0.34 \\
    4 & 3 & y & 2.1670 &  0.0009   & 0.64 \\
    5 & 3 & y & 2.1657 &  0.0020   & 0.49 \\
\end{tabular}
\end{center}
\label{tab:fitres}
\end{table}
\begin{table}[htbp]
\caption{Comparison of recent results for the dynamic exponent $z$.
Numerical errors are in parentheses.
}
\begin{center}
\begin{tabular}{||l|d|l||}
Reference                               & Year & Value             \\
\hline
Present work                            & 1996 &  2.1665  (12)    \\
Li et al.~\cite{LiSZ}                   & 1995 &  2.1337  (41)    \\
Linke et al.~\cite{Leal}                & 1995 &  2.160    (5)    \\
Grassberger~\cite{Gras}                 & 1995 &  2.172    (6)    \\
Wang et al.~\cite{Wetal}                & 1995 &  2.16     (4)    \\
Baker and Erpenback~\cite{BakerErp}	& 1994 &  2.17     (1)    \\
Ito~\cite{Ito}                          & 1993 &  2.165   (10)    \\
Dammann and Reger~\cite{DamReg}         & 1993 &  2.183    (5)    \\
Matz et al.~\cite{MHJ}                  & 1993 &  2.35     (5)    \\
M\"unkel et al.~\cite{MHAG}             & 1993 &  2.21     (3)    \\
Stauffer~\cite{Stauf1}                  & 1993 &  2.06     (2)    \\
\end{tabular}
\end{center}
\label{tab:compare}
\end{table}
\end{document}